\documentclass[graybox]{svmult}
\usepackage[utf8]{inputenc}
\usepackage{hyperref}
\usepackage{fullpage}
\usepackage{enumitem} 
\usepackage[normalem]{ulem}
\usepackage{soul}
\usepackage{upgreek}
\usepackage{boxhandler}
\usepackage[bottom]{footmisc}
\usepackage{newtxmath}       

\usepackage{graphicx}
\graphicspath{ {images/} }
\usepackage{float}
\usepackage{morefloats} 
\usepackage{wrapfig}
\usepackage{caption}
\usepackage{subcaption}
\usepackage{xcolor}

\usepackage{mathtools}
\usepackage{mathrsfs}
\usepackage{slashed}
\usepackage{cancel}
\usepackage{braket} 

\newcommand{\av}[1]{\langle #1 \rangle}
\newcommand{\mink}{\mathbb M} 
\newcommand{\be}{\begin{equation}}
\newcommand{\ee}{\end{equation}}
\newcommand{\ba}{\begin{align}}
\newcommand{\ea}{\end{align}}
\newcommand{\bfig}{\begin{figure}}
\newcommand{\efig}{\end{figure}}
\newcommand{\bsfig}{\begin{subfigure}}
\newcommand{\esfig}{\end{subfigure}}
\newcommand{\kgnorm}[1]{( #1 )_{\mathrm{KG}}}
\newcommand{\hD}{\widehat{\Delta}} 
\newcommand{\uksj}{s_{\mathbf k}}
\newcommand{\uksjn}{\tilde{s}_{\mathbf k}}
\newcommand{\ukpsj}{s_{\mathbf k'}}
\newcommand{\vkpjp}{\tilde{s}_{\mathbf k}^+}
\newcommand{\vkpjm}{\tilde{s}_{\mathbf k}^-}
\newcommand{\uqkg}{u_{\mathbf q}}
\newcommand{\uqpkg}{u_{\mathbf q'}}
\newcommand{\Aqk}{A_{\mathbf{qk}}}
\newcommand{\Bqk}{B_{\mathbf{qk}}}
\newcommand{\hP}{\widehat{\Phi}}
\newcommand{\hB}{\widehat{\Box}}
\newcommand{\wsj}{W_{\mathrm{SJ}}}
\newcommand{\kr}{\mathrm{Ker}} 
\newcommand{\im}{\mathrm{Im}} 
\newcommand{\ha}{\hat{\mathbf{a}}}
\newcommand{\hb}{\hat{\mathbf{b}}}
\newcommand{\hbk}{\hb_{\mathbf {k}}}
\newcommand{\hbkd}{{\hb_{\mathbf {k}}^\dagger}}
\newcommand{\haq}{\ha_{\mathbf {q}}}
\newcommand{\haqd}{{\ha_{\mathbf {q}}^\dagger}}
\newcommand{\haqp}{\ha_{\mathbf {q'}}}
\newcommand{\haqpd}{{\ha_{\mathbf {q'}}^\dagger}}
\newcommand{\lk}{\lambda_{\mathbf{k}}}
\newcommand{\tre}{\text{Re}}
\newcommand{\cL}{\mathcal{L}}
\newcommand{\Wmink}{W_{\mathrm{mink}}}
\newcommand{\phm}{m_p}

\begin{document}

\title*{Quantum Field Theory On Causal Sets}

\author{Nomaan X}

\institute{Nomaan X \at Department of Mathematics and Statistics, University of New Brunswick, Fredericton NB, Canada,\\ \email{nomaan.math@unb.ca}}

\maketitle

\abstract{We give a broad overview of a construction of a theory for matter on fixed causal set backgrounds. We introduce the Sorkin-Johnston formalism for a free (real) scalar field theory that is applicable to regions of continuum spacetimes as well as to causal sets. We show examples in the causal set, starting from the construction of Green functions to obtaining unique two-point functions using this formalism. We also mention other approaches that have been explored in constructing dynamics for matter on causal sets, including ideas for interacting theories and fermions.}


\section{Introduction}
\label{sec:intro}

The main motivation for wanting to describe QFT on a discrete spacetime background is to avoid divergences\footnote{Lattice field theory is the most straightforward but less foundational implementation of this idea with the advantage of being amenable to numerical/computational methods.}. Traditionally, this was handled by a variety of renormalization techniques which, initially introduced as technical tools to obtain sensible results, are now understood in the sense of Wilsonian renormalization as ways of obtaining an effective field theory. The idea being that since we don't know the full UV complete theory, we're always dealing with effective descriptions involving energy dependent parameters. These effective descriptions stop making sense beyond certain energy scales. This approach has seen stunning success as far as high energy physics phenomenology goes. It has also become the standard way of studying condensed matter systems.

In the study of quantum gravity however, this approach has severe limitations. Firstly, by construction it throws away information on the deep UV regime. Secondly, while the renormalization group flow description of couplings in a theory can indicate the breakdown of effective theories, it will not signal the appearance of radically new physics that may arise. Ideas such as non-locality, stochastic behaviour of matter (or gravity), causality violation or the breakdown of notions tied to the continuum are not trivial extensions of local, renormalizable, continuum based effective field theories.

The interest in studying QFT on causal sets is not only a way around these limitations but is in itself an important ingredient in thinking of causal sets as fundamental to quantum gravity. It is foundational and bottom-up in the same sense as Boltzmann's statistical mechanics was in describing standard macroscopic thermodynamics of his time. Even putting aside the question of whether causal sets are in fact fundamental or not, we might be interested in studying QFT on a discrete background which, unlike say a simple hyper-cubic lattice, shares basic properties of continuum spacetime (e.g. Lorentzian signature and Lorentz-invariance). Effects of spacetime discreteness from the causal set may show up at larger scales in observations and/or experiments; at the very least, it will help us constrain the discreteness parameters. 

Various approaches to studying QFT on causal sets have been considered and below we provide a brief summary - 

\begin{itemize}
    \item \textit{Matter as a causal set}: These are inspired from Wheeler's geometrodymanics where, field configurations are encoded in patterns arising from the causal structure i.e., matter arises from spacetime. An example has been proposed here \cite{stanley1986enumerative}. Another possibility is to model matter \`ala Kaluza-Klein, where we start with a higher dimensional causal set dynamics and try to split this into a lower dimensional causal set part along with a matter part. This is yet to be implemented. 
    \item \textit{Matter on a causal set}: This is along the lines of standard QFT where dynamics of matter are described on a background spacetime. There have been multiple approaches to this - 
    \begin{enumerate}
        \item Scalar field theory in histories form: Histories based formulations of quantum theories are regarded as more satisfactory than operator formulations for various reasons. Such an approach for a free scalar field on causal sets is based on the {\it Decoherence functional} $D(\zeta,\bar{\zeta})$ that maps pairs of spacetime histories\footnote{A pair is called a {\it Schwinger history} in honor of the Schwinger-Keldysh version of the path integral.} of the scalar field to complex numbers. The dynamics must then described in terms of $D$ and a quantum measure obtained from it. Suggestions for generalizations to interacting theory have also been proposed \cite{Sorkin:2011pn}.   
        \item Green functions and the Sorkin-Johnston formalism: In this approach, the classical equations of motion are bypassed and we directly identify a retarded Green function for the scalar field. Then we use the SJ formalism to construct a unique Wightman function for the theory \cite{Johnston:2010su,Sorkin:2017fcp}. This approach has been studied the most, not just in the context of causal sets \cite{Johnston:2010su,Surya:2018byh} but also in the continuum \cite{Afshordi:2012ez,Sorkin:2017fcp,Mathur2019}, where it can be used as an alternative to canonical quantization. Importantly, it provides a way around the problem of the choice of a vacuum in arbitrary spacetime regions.
    	
        Recently, an alternate way of defining a vacuum state using the notion of geometric quantization instead of using the SJ axioms (see below) has been proposed\cite{Hawkins2022}. This method can be applied to symplectic manifolds with a Riemannian metric. In the case of causal sets this means applying the procedure to a symplectic vector space with an inner product, which results in a state identical to the SJ state. 
    	
        \item d'Alembertians: A scalar field on continuum spacetime satisfies the Klein-Gordon equation. A natural step in discretization is therefore to find the causal set analog of the d'Alembertian operator. One way to do this is to identify a Green function and then invert it. This was tried in regions of $\mink^2$ \cite{daughton1993the,salgado2008toward} and found to be a good approximation to the continuum d’Alembertian for fields which vary slowly on the discreteness scale and are zero on the boundary of the region considered.\\
        Another, better studied way is to construct a d'Alembertian operator in a way analogous to the usual second order differential operator, as a difference operator using nearest neighbours. However, due to the Lorentzian nature of the metric it is not obvious how to identify ``nearest neighbours''. A way around this was the proposal to use a past-layered decomposition of the causal set and then sum over these layers with appropriate coefficients \cite{sorkin2009does,Benincasa:2010ac}. A class of such d'Alembertians has now been constructed for different geometries and dimensions \cite{dowker2013causal,aslanbeigi2014generalized}.\\
        It remains an open question as to how these d'Alembertians can be used to define a full field theory on the causal set. 
        \item Field Lagrangians: This approach is based on re-expressing the classical field Lagrangian in terms of the causal structure, the volume element and the proper time i.e., quantities derived from the causal set \cite{sverdlov2009gravity}. While this has the advantage of being amenable to gauge fields and spinors, it is a top-down approach that starts from the continuum and leads to a somewhat complicated causal set expression. Instead, it would be appealing to start from things defined on the causal set and then take the continuum limit as a cross-check.       
    \end{enumerate}
\end{itemize}

We note that while the motivations for constructing QFT on causal sets come from quantum gravity, such a construction, as a UV complete theory would require evaluating the full path integral over all causal sets and field configurations. This is beyond our current understanding of the subject. Instead, we build QFT on sprinkled causal sets i.e., causal sets approximated by fixed regions of continuum spacetime. This way, we can focus on aspects of quantum fields on causal sets without the complications arising from dynamics of causal sets themselves. This corresponds to working at a mesoscale that is not quite the Planck (or discreteness) scale but incorporates the effects of discreteness. Hence, we probe higher energies than QFT on curved spacetime, which might be phenomenologically relevant.    

\section{The Sorkin-Johnston Method} 

The method we describe here is for the construction of a theory of a free (real) scalar field. We also note that this construction can be carried out rigorously in the continuum as well \cite{Afshordi:2012jf,Sorkin:2017fcp,Fewster:2018ltq}. Here we review the continuum construction and then restrict ourselves to the causal set where the construction is greatly simplified.

We will proceed in an unorthodox manner that is more suited to the causal set - rather than start with the field equations\footnote{While there are algebraic conditions that can potentially be used as equations of motion on the causal set, they are not analogous to the dynamical equations that we are used to in standard field theory.}, the construction will be based on the retarded Green function of the field. The two-point correlation function or the {\it Wightman function} can then be derived from this Green function
\be
G \longrightarrow \Delta \longrightarrow W.
\ee
where $\Delta$ is called the \textit{Pauli-Jordan function} and we define it below.
Such an approach builds in state information from the very beginning, in the form of the Wightman function, $W(x,x')$. 

We begin our discussion with the first step in the construction, the identification of appropriate Green functions on the causal set. In general, there is no known way to obtain a Green function starting from the causal set. The examples we discuss below are some known and interesting cases where the Green function on the causal set can be motivated from the knowledge of the corresponding continuum Green function. We will also see that there are usually multiple ways to do this and we expect that they converge in the continuum limit. Towards the end of the following section we briefly mention alternate proposals for obtaining Green functions. 

\subsection{Green Functions}

Before we discuss examples of constructing Green functions, we note an important connection between the massless and massive Green functions. Consider the massless scalar retarded Green function $G_0(x,x')$ on a
globally hyperbolic  $d$ dimensional spacetime $(M, g)$:
\be
\Box_x G_0(x,x')=- \frac{1}{\sqrt{-g(x')} }\delta(x-x')\,.
\ee
The massive retarded Green function, $G_m$, satisfies
\be
(\Box_x - m^2) G_m(x,x')= - \frac{1}{\sqrt{-g(x)} }\delta(x-x')\,,
\ee
and can be written as a formal expansion
\be \label{eq:conv}
G_m=G_0 - m^2\,G_0*G_0 + m^4 \,G_0*G_0*G_0+ \ldots = \sum_{k=0}^\infty(-m^2)^k \underbrace{G_0 * G_0* \ldots G_0}_{k+1}
\ee
where
\be
(A\ast B)(x,x')\equiv \int d^dx_1 \sqrt{-g(x_1)} A(x,x_1) B(x_1,x')\,.
\ee
If $G_0(x,x')$ is retarded then so is $G_m(x,x')$.

Consider the more general case of a scalar theory in curved spacetime. The Green function satisfies
\be
(\Box_g - m^2 -\xi R)G_{m,\xi}(x,x')=\frac{1}{\sqrt{-g(x)} }\delta(x-x')\,.
\ee
$G_{m,\xi}(x,x')$ can be obtained from $G_{0,\xi}(x,x')$ using the same series expansion Eq.\eqref{eq:conv}:
\be
G_{m,\xi}= \sum_{k=0}^\infty(-m^2)^k \underbrace{G_{0,\xi} * G_{0,\xi}* \ldots G_{0,\xi}}_{k+1} \,.
\ee

Further, in the special case when $R$ is a constant, the $\xi R$ term just modifies the mass and  $G_{m,\xi}(x,x')$  can be obtained from the massless minimally coupled (MMC) Green  function $G_{0,0}(x,x')$ with $m^2$ replaced by $m^2+\xi R$.
In general, for constant $R$, we can relate the two Green functions 
\be \label{eq:convxi}
G_{m', \xi'}= \sum_{k=0}^\infty(-{m'}^2 - \xi' R +m^2 + \xi R)^k \underbrace{G_{m,\xi} * G_{m,\xi}*
	\ldots G_{m,\xi}}_{k+1} \, , 
\ee
for any $(m,\xi)$, $(m',\xi')$.

Therefore, if we have the massless (or MMC) retarded Green function, we can write down a formal series for the massive retarded Green function. Note that, in the $d=2$ case, this formal series contains IR divergences in each term due to the presence of $G_0$\cite{Aste:2007ii}. However, on causal sets the expression is well defined since we are always working with matrices and finite sums.

If we have a massless (or MMC) retarded Green function analogue\footnote{We avoid writing $\xi$ in the causal set expressions, however, when appropriate, $\xi$ will modify the mass as discussed for the continuum case.}, $K_0(x,x')$, on a causal set with sprinkling density $\rho$ in a volume $V$ of $d$-dimensional spacetime, we can propose a massive retarded Green function $K_m(x,x')$ via the replacement
\be
\int \sqrt{-g(x)}\,d^d x \rightarrow  \rho^{-1} \sum_{\textrm{causal set elements} }\,,
\ee
leading to
\be \label{eq:convK}
K_m = \sum_{k=0}^{\infty}\left(-\frac{m^2}{\rho}\right)^k \underbrace{K_0 . K_0. \ldots K_0}_{k+1}
\ee
where now the convolutions become matrix products. The series terminates and is well-defined for each pair $x$ and $x'$ because, as we will see, the matrices involved in the products are nilpotent. 

The key to the above construction of a massive Green function is knowing the massless one. We can repeat it on causal sets if we can find the appropriate massless retarded Green function analogues for causal sets sprinkled into general curved spacetimes.

We now show a few examples of explicit construction of Green functions\cite{Johnston:2010su,nomaan2017scalar}.

\subsubsection{$d=2$ Minkowski}

The massless retarded Green  function in $d=2$ Minkowski spacetime $\mink^2$ is
\be \label{eq:massless2dcont}
G^{(2)}_0(x,x')=  \frac{1}{2}\theta(x_0-x'_0)\theta(\tau^2(x,x'))
\ee
where $\tau$ is the proper time and $\theta$ is the Heaviside step function.

For any causal set $\mathcal{C}$, we can define a causal matrix $C_0(x,x')$ as 
  \[  C_0(x,x'):=                                        
\left\{                                                          
\begin{array}{ll}
	1  & \mbox{if } x' \prec x \\
	0 & \mbox{} \text{otherwise}
\end{array}.
\right.
\]
The Poisson sprinkling gives a random variable, which we also call $C_0(x,x')$, for every two points, $x$ and $x'$ on that $\mathcal{C}$. It was shown\footnote{This is also intuitively clear from the definition of the causal matrix.} \cite{daughton1993the} that the average value of this variable is
\be
\av{C_0(x,x')} =  2 G^{(2)}_0(x,x')\,.
\ee
This suggests that the causal matrix as the analogue Green function in this case
\be \label{eq:massless2dcs}
K^{(2)}_0(x,x')\equiv  \frac{1}{2}C_0(x,x').
\ee
Then a massive Green function $K^{(2)}_m(x,x')$ on $\mathcal{C}$ can be defined using this and Eq.\eqref{eq:convK} as
\be
K^{(2)}_m(x,x') =  \sum\limits_{k=0}^\infty  \biggl(- \frac{m^2}{\rho}\biggr)^{k}\biggl(\frac{1}{2}\biggr)^{k+1} C_k(x,x')\,,
\ee
where $C_k$'s are called $k$-chains and are powers of the causal matrix - $C_k(x,x') = \underbrace{C_0 \cdot C_0\cdot \ldots C_0}_{k+1} (x,x')$.

It can be shown that the average value of the corresponding random variable, for any sprinkling density, is equal to the continuum massive Green function
\begin{align}
	\av{K^{(2)}_m(x,x')} & = G^{(2)}_m(x,x')\,.
\end{align}

Another way to interpret $K^{(2)}_m(x,x')$ is in terms of {\it hop} and {\it stop} weights, $a$ and $b$ respectively \cite{Johnston2008particle}:
\be
K^{(2)}_m(x,x') =  \sum\limits_{k=0}^\infty  a^{k+1} b^k C_k(x,x').
\ee
This form is interpreted as a sum over all chains between $x$ and $x'$: for each $k$-chain the hop between two successive elements has weight $a$ and the stop at each intervening element between $x$ and $x'$ has weight $b$. We see that the weight $a=1/2$ is associated to each factor of $K^{(2)}_0$ (from the relationship between $K^{(2)}_0$ and the causal matrix) and the weight $b=-m^2/\rho$ to each convolution. We will see below that in certain spacetimes with curvature, manipulating these weights appropriately will give us the right Green function. 

\subsubsection{$d=4$ Minkowski}

In $d=4$ Minkowski spacetime, $\mink^4$, the retarded Green  function for the massless field is
\be \label{eq:massless4dcont}
G^{(4)}_0(x,x')= \frac{1}{2\pi}\theta(x_0-x'_0) \delta( \tau^2(x,x'))\,,
\ee

The causal set analogue is proportional to the link matrix defined as
\[L_0(x,x'):=                                        
\left\{                                                          
\begin{array}{ll}
	1  & \mbox{if } x' \prec x \,\text{and}\, |(x,x')|=0\\
	0 & \mbox{} \text{otherwise}
\end{array},
\right.
\]

The average value of the corresponding random variable $L_0(x,x')$ in a Poisson sprinkling of density $\rho$ is
\be \label{eq:linkexp4d}
\av{L_0(x,x')}=\theta(x_0-x'_0) \theta( \tau^2(x,x'))\exp(-\rho V(x,x')),
\ee
where $V(x,x')$ is the volume of the spacetime interval\footnote{Such an interval is called a {\it Causal diamond} or {\it Alexandrov interval}.} $J^-(x) \cap J^+(x')$. Using $V(x,x')= \frac{\pi}{24}\tau^4(x,x')$, it can be shown that
\begin{align}
	\lim_{\rho \rightarrow \infty} {\sqrt{\frac{\rho}{6}}\av{L_0(x,x')}}& = 2 \pi G^{(4)}_0(x,x')\,.
\end{align}
This suggests that we pick the massless Green function in this case as 
\be \label{eq:massless4dcs}
K^{(4)}_0(x,x')= \frac{1}{2 \pi} \sqrt{\frac{\rho}{6}} L_0(x,x')\,.
\ee
\bfig[H] 
\begin{center} 
	\includegraphics[width = .56 \textwidth]{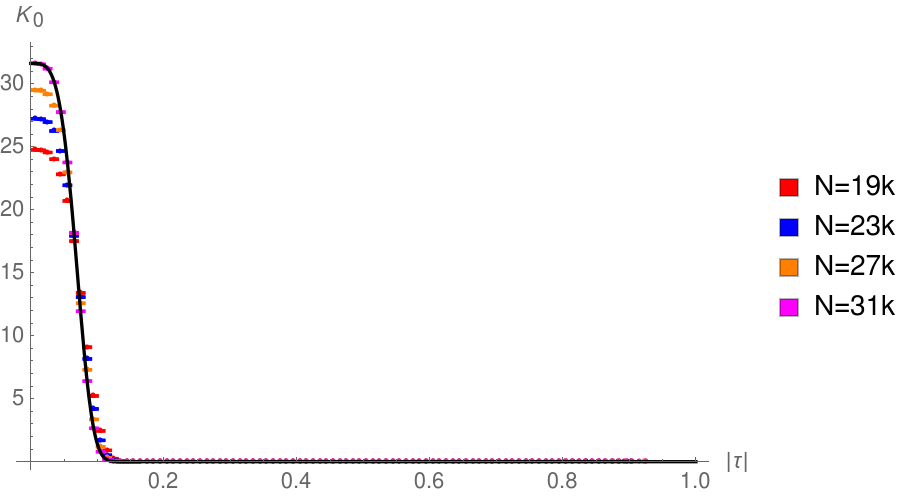}
	\caption{The binned and averaged plot for $K_0$ vs. $|\tau|$ (timelike distance) as $N$ is varied. The black curve represents the average value for $N=31k$.}
 \label{4dcdgreen}
\end{center} 
\efig
The relationship with the continuum Green function is not as direct here because it is only in the continuum limit as $\rho\rightarrow \infty$ that the average value of $K^{(4)}_0$ equals the continuum $G^{(4)}_0$. In figure \ref{4dcdgreen} we plot binned and averaged values for the causal set retarded Green function Eq.\eqref{eq:massless4dcs} along with its average value at finite density obtained from Eq.\eqref{eq:linkexp4d} in a causal diamond of height unity. The corresponding continuum Green function Eq.\eqref{eq:massless4dcont} has a delta function on the lightcone and is therefore infinitely sharply peaked there. While this is not the case in the causal set, the discrepancy grows smaller as the density is increased.

We use this $K^{(4)}_0$  to construct a massive Green function $K^{(4)}_m(x,x')$ via Eq.\eqref{eq:convK} as before
\be \label{eq:discreteGF4d}
K^{(4)}_m(x,x') =  \sum\limits_{k=0}^\infty  \biggl(-\frac{m^2}{\rho}\biggr)^{k}\biggl(\frac{1}{2\pi}\sqrt{\frac{\rho}{6}}\biggr)^{k+1} L_k(x,x')\,.
\ee

For the corresponding random variable one can show that
\begin{align}
	\lim_{\rho \rightarrow \infty} \av{K^{(4)}_m(x,x')} & = G^{(4)}_m(x,x')\,.
\end{align}

The hop-stop weights can be read off from Eq.\eqref{eq:discreteGF4d} as
$a=\frac{1}{2 \pi} \sqrt{\frac{\rho}{6}} $ and $b=-\frac{m^2}{\rho}$, respectively.

\subsubsection{$d=4$ dS and adS}

In $d=4$ for conformally flat spacetimes $g_{ab}=\Omega^2(x)\eta_{ab}$ the conformally coupled massless Green function is related to that in $\mink^4$ by  
\be 
G_{0,\xi_c}(x,x')=\Omega^{-1}(x) G_0^{F}(x,x')\Omega^{-1}(x'),  
\ee 
where $\xi_c = \frac{1}{6}$ and  $G_0^F(x,x')$ is the retarded massless Green function in $\mink^4$. 
Further, if we have constant scalar curvature, the massive Green function for arbitrary $\xi$ can be obtained from $G_{0,\xi_c}(x,x')$  using Eq.\eqref{eq:convxi} because $\xi R$ is then a mass term.

In global dS spacetime and in the conformally flat patch of adS spacetime, it has been shown that \cite{nomaan2017scalar}
\be
\lim_{\rho \rightarrow \infty} \frac{1}{2\pi}\sqrt{\frac{\rho}{6}} \av{L_0(x,x')}=G_{0,\xi_c}(x,x'). 
\ee 
Therefore, we can construct the massive retarded Green function for arbitrary coupling from this along the same lines as before
\be
K^{(4)}_{m,\xi}(x,x') \equiv   \sum\limits_{k=0}^\infty  a^{k}b^{k+1} L_k(x,x').
\ee
with $a=\frac{1}{2\pi}\sqrt{\frac{\rho}{6}}$ and $b = -\frac{m^2+ (\xi -\frac{1}{6})R}{\rho}$.

\subsubsection{Riemann Normal Neighbourhoods}

In more general spacetimes, it is a highly non-trivial task to find the average values of the random variable defined through the causal or link matrices. Due to this, it is not possible to prove if these random variables have the right continuum limit that we desire. However, there are still situations where we can make a more quantitative guess about the causal set Green function. 

For example, every $d=2$  spacetime is locally conformally flat. If the spacetime is topologically trivial then the MMC Green function equals the flat spacetime Green function Eq.\eqref{eq:massless2dcont}. Therefore, it seems reasonable to expect, at least locally, that on causal sets sprinkled into such spacetimes, the massless minimally coupled causal set Green function, $K^{(2)}_{0,0}(x,x')$, is the flat one given by Eq.\eqref{eq:massless2dcs} and therefore for regions where $R$ is approximately constant, that $K^{(2)}_{m,\xi}(x,x')$ is given by:
\be
K^{(2)}_{m,\xi}(x,x') =  \sum\limits_{k=0}^\infty  \biggl(- \frac{m^2+\xi R}{\rho}\biggr)^{k}\biggl(\frac{1}{2}\biggr)^{k+1} C_k(x,x')\,,
\ee
The argument that the average value over sprinklings of the corresponding random variable will be the correct continuum Green function proceeds exactly as in the flat case.
This is still a heuristic argument and cannot be made more concrete because $\av{C_k(x,x')}$ is not known. The best we can do is show that the average value gives the right Green function in a Riemann normal neighbourhood (RNN), in the continuum limit to leading order. Indeed, we find that when $m^2\gg\xi R$:
\begin{align}
G^{(2)}_{m,\xi}(x,x') &\approx \theta(x_0) \theta(\tau^2)\left[  \dfrac{1}{2}J_{0}(m\tau) +\dfrac{R(x')\tau^2}{48}J_{2}(m\tau) - \frac{\xi R(x')\tau}{4m}J_1(m\tau)\right]\nonumber\\
&=\lim_{\rho\rightarrow\infty}\av{K^{(2)}_{m,\xi}(x,x')}
\end{align}

One can prove a similar result in an RNN in $d=4$ starting with the link matrix with a further constraint that the average value matches only when $R_{ab}(x')\propto g_{ab}(x')$ i.e., for Einstein spaces. We refer the reader to \cite{nomaan2017scalar} for more detail on these calculations.

We emphasize that in the above constructions, the apriori knowledge of the continuum Green function was used to first propose an analogue on the causal set and then check if it gives the right continuum limit.

Before ending this section we also mention two other, potentially more fundamental, proposals for constructing a Green function - 
\begin{itemize}
    \item \textit{using d'Alembertians}: In the continuum, we expect there to be an inverse relation between the d'Alembertian and the Green function (with appropriate boundry conditions). It is reasonable to expect that such a relation might exist for the analogous objects in the causal set. As mentioned in the introduction, there are well studied proposals for constructing d'Alembertians on causal sets. These d'Alembertians are, by construction, retarted i.e., they are built out of layers in the past of a given causal set element, one might expect that their inverse\footnote{In the causal set, this is a simple matrix inversion.} should therefore define some sort of retarted Green function. The important question is whether this is the retarted Green function we want i.e., the one that gives the right continuum limit. Such a "sweety-salty" duality has been shown in a limited context \cite{johnston2015correction} of $\mink^4$ but might be more general.
    \item \textit{using preferred past}: A more recent proposal for constructing a Green function augments the causal set\footnote{Another such attempt was made earlier on by introducing a slicing and using it to define discrete d'Alembertians \cite{Foster_2004}.} with a \textit{preferred past structure} in order to first define a d'Alembertian and then a retarted Green function. This is motived from ideas in local algebraic quantum field theory \cite{Dable-Heath:2019sej}. The preferred past structure is a map $\Lambda$ such that the preferred past $\Lambda(p)$ of a point $p$ is a point of rank\footnote{Rank in this context means the minimum number of links in a path between 2 points.} 2 in the past of $p$. In general, this structure can be chosen in different ways and it is not clear which is the most appropriate choice. An example of theory construction for a regular diamond lattice using this method is shown in \cite{Dable-Heath:2019sej}.                                                                                                                 
\end{itemize}

\subsection{SJ in the continuum}

The next step in the construction is the choice of a state. Consider a region with finite volume $V$ in $(M,g)$. For a free scalar field in this region the Klein Gordon (KG) equation is  
\be 
	\biggl(\hB- m^2\biggr) \phi=0,
\ee 
where $\hB \equiv g^{ab}\nabla_a \nabla_b$, and the effective mass $m^2=\phm^2 +\xi R$, where $\phm$ is the physical
mass, $R$ is the scalar curvature and $\xi$ is the coupling.   
Let $\{\uqkg\}$ be a complete set of modes satisfying the KG equation in $(M,g)$ and orthonormal with respect to the KG inner product
\be
	\kgnorm{f,g}=\int_\Sigma (f^*\nabla_a g-g^*\nabla_a f) dS^a,  
\ee 
where $\Sigma$ is a Cauchy hypersurface in $(M,g)$. The field operator corresponding to the classical field can be expressed as a mode expansion with respect to this set    
\be
	\hP(x) \equiv \sum_{\mathbf{q}}\haq \uqkg(x) +\haqd\uqkg^*(x), 
\ee
with $\haq,\,\haqd$
satisfying the commutation relations
\be
	[\haq,\haqpd]=\delta_{\mathbf{q}\mathbf{q}'},\quad [\haq,\haqp]=0\text{,}\quad [\haqd,\haqpd]=0. 
 \ee 
The covariant commutation relations for the scalar field operator are given by the {\it Peierls bracket}
\be
	[\hP(x),\hP(x')] = i\Delta(x,x'), 
\ee 
where the Pauli-Jordan (PJ) function $\Delta(x,x')$ is given by
\be  
	\Delta(x,x')  \equiv G_R(x,x')-G_A(x,x'),
\ee 
with $G_{R,A}(x,x')$ being the retarded and advanced Green functions, respectively. This can also be written in terms of the modes $\{\uqkg\}$ using the mode expansion and the commutation relations  
\be
	i \Delta(x,x')=\sum_{\mathbf{q}}\uqkg(x)\uqkg^*(x')-\uqkg^*(x)\uqkg(x'), 
\ee
and the two-point function or the state/vacuum\footnote{In the usual language of Fock spaces, the vacuum would be $\haq\,\ket{0}=0$. } associated with these modes is defined as the positive part of the above expansion
\be
	W(x,x') \equiv \sum_{\mathbf{q}}\uqkg(x)\uqkg^*(x'). 
\ee
The initial choice of modes is usually motivated by a choice of observer or the symmetries of the background spacetime. The SJ state or equivalently the SJ modes are constructed directly from $i \hD$ in any given {\it finite} spacetime region, do not require a choice of observer and are thus unique. The question of physical interpretation of the SJ state is more nuanced and has been studied in some special cases \cite{Afshordi:2012jf}. 

To construct the SJ vacuum from the PJ function, it is elevated to an integral operator as follows
\be
	i \hD \circ f \equiv i \int_{V} \Delta(x,x')  f(x') dV_{x'}
\ee 
which acts on $\cL^2$ functions in $V$ and where  
\be
	\braket{f,g}=\int_V dV_x\,f^*(x)\,g(x) 
\ee  
is the $\cL^2$ inner product.  
Since $\Delta(x,x')$ is  antisymmetric in its arguments, $i\hD$ is Hermitian on the space of $\cL^2$ functions in $V$. Its non-zero eigenvalues, given by  
\be
	i \hD \circ \uksjn(x) =\int_{V} dV_{x'}\,i\Delta(x,x')\uksjn(x')=\lk \uksjn(x)
\ee 
therefore come in pairs $(\lk, -\lk)$, corresponding to the eigenfunctions $(\vkpjp, \vkpjm)$ where $\vkpjm=(\vkpjp)^*$.\footnote{We adopt the notation that the $\tilde{s}_k$ are the un-normalised (with respect to the $\mathcal{L}^2$ norm) SJ eigenfunctions,  whereas the $s_k$ without the tilde are the normalised SJ eigenfunctions.} One therefore has an {\it intrinsic} and {\it coordinate/observer independent} separation of positive and negative eigenmodes of $i\hD$.

The central idea of the construction is the following observation \cite{Wald:1995yp,Sorkin:2017fcp} 
\be \label{eq:kerim}
	\kr (\hB -\phm^2) = \overline{\im (\hD)}, 
\ee
where the operators are defined in $V$\footnote{In a spacetime of constant scalar curvature, $m$ defined above is constant, and hence this result continues to hold when $\phm$ is replaced by $m$.}. This means that the eigenvectors in the image of $i \hD$ (i.e., excluding those in $\kr (i\hD)$) span the full solution space of the KG operator. Along with our earlier result that these eigenvectors have a separation into those with positive eigenvalues and those with negative eigenvalues, we get a unique decomposition of the field operator
\be
	\hP(x)=\sum_{\mathbf{k}}\hbk\uksj(x)+\hbkd \uksj^*(x),
\ee
and the SJ vacuum state is defined as 
\be 
	\hbk\ket{0_{SJ}}=0 \quad \forall\,\mathbf{k},
\ee where  
\be
	\uksj = \sqrt{\lk} \vkpjp   
\ee  
are the normalised {\sl SJ modes} which form an orthonormal set in $\overline{\im(i\hD)}$ with respect to the $\cL^2$ norm
\begin{align}
	\braket{\uksj,\ukpsj}&={\lk}\delta_{\mathbf{k}\mathbf{k}'} \nonumber \\ 
	\braket{\uksj^*,\ukpsj}&=0.
\end{align}
Using the spectral decomposition 
\be 
	i \Delta(x,x') = \sum_{\mathbf k}  \uksj(x) \uksj^*(x') -\uksj^*(x) \uksj(x'), 
\ee
the SJ two-point function in $V$ is the positive part of $i\hD$ 
\be 
	\wsj(x,x') \equiv  \sum_{\mathbf k} \uksj(x) \uksj^*(x'). 
\ee 
If $\wsj(x,x')$ remains well-defined as the IR cutoff (i.e., $V$) is taken to infinity, this defines the SJ vacuum in the full spacetime $(M,g)$.

Alternately, The SJ two point function is uniquely defined most generally by the following conditions \cite{Sorkin:2017fcp} 
\begin{align}
	&i\Delta(x,x') = \wsj(x,x') - \wsj(x',x),  \nonumber\\
	&\int_{V} dV' \int_{V} dV f^*(x') \wsj(x',x) f(x)  \geq 0,\quad(\text{Positive Semidefinite}) \nonumber \\ 
	&\int_V dV' \wsj(x,x') \wsj^*(x',x'') = 0,\quad(\text{Ground state or Purity}) 
\end{align}
The first condition follows from the definition of the Wightman function. The motivation for the second condition can be seen as follows - given a state vector $|0\rangle$ in some Hilbert space the above result can be derived as a theorem that follows immediately from the positivity of $||\psi||^2=\av{\psi|\psi}$ where $|\psi\rangle=\int dV(x)f(x)\,\hat{\phi}\,|0\rangle$. The utility of the final condition is not as obvious, we just mention that this condition helps in picking out, from a set of solutions, those that match the notion of a ground state when such a notion is available. For example, it has been shown that the SJ vacuum will coincide with the minimum energy vacuum in stationary spacetimes\cite{Afshordi:2012jf,Afshordi:2012ez}. In other cases this condition implies that the entanglement entropy associated with $\wsj$ vanishes i.e., the SJ state is always pure. We refer the reader to \cite{Sorkin:2017fcp} for a simple example of these conditions in action.   

A third way to obtain the SJ modes is via a mode comparison using Bogoliubov coefficients \cite{Afshordi:2012jf}. Given the equality in Eq.\eqref{eq:kerim} between $\overline{\im(\hD)}$ and the KG solution space, there must exist a transformation between the KG modes $\{ \uqkg\}$ in $V$ and the SJ modes $\{\uksj\}$, even if the former are not orthonormal with respect to the $\cL^2$ inner product. Namely, we can write
\be
	\uksj(x)=\sum_{\mathbf{q}}\uqkg(x)\Aqk+\uqkg^*(x)\Bqk,
\ee
where $\Aqk=\kgnorm{\uqkg,\uksj}$, $\Bqk =\kgnorm{\uqkg^*,\uksj}$ and they satisfy the constraints  
\begin{align}
	\sum_{\mathbf{q}}\mathbf{A_{qk'}}\Aqk^*-\mathbf{B_{qk'}}\Bqk^*&=\delta_{\mathbf{k}\mathbf{k}'}\nonumber\\
	\sum_{\mathbf{q}}\mathbf{B_{qk'}} \Aqk -\mathbf{A_{qk'}} \Bqk &=0.
\end{align}
Further, if the KG modes themselves satisfy the $\cL^2$ orthonormality condition 
\be
	\langle \uqkg,\uqpkg\rangle=\delta_{qq'},\indent \langle \uqkg^*,\uqpkg \rangle=0,  
\ee
the constraints simplify considerably.

It is important to note that the above calculations are limited to finite $V$. There are subtleties in identifying $\kr (\hB -m^2)$ in $V$, starting from the solutions in full spacetime.  

An important question is whether the limits involved in these approaches commute. In the first two approaches we define the SJ vacuum directly in finite $V$ and only take the limit $V\rightarrow\infty$ at the end, if possible, whereas in the mode comparison approach we might have to take the limit for the comparison.
A case in point is the 2d causal diamond in Minkowski spacetime, where, in order to compare with the IR limit, $W(x,x')$ was studied in a small region in the interior of the larger diamond, which to leading order was found to have the form of the (IR-regulated) 2d  Minkowski vacuum \cite{Afshordi:2012ez}. Similar considerations come up in \cite{Afshordi:2012jf,Aslanbeigi:2013fga} when using the Bogoliubov method.

Before moving on to the construction in the causal set, we mention the Hadamard condition, which is related to the UV behavior of two-point functions. It has been shown \cite{fewster2012on,Brum:2013bia} that the SJ state in the continuum is not Hadamard. Further, one can obtain a Hadamard state by introducing a smoothening function in the definition of the PJ operator, this has been tested in static and cosmological spacetimes \cite{Brum:2013bia}. However, the introduction of such a smoothening function introduces non-uniqueness into the construction. Whether a unique, physically motivated choice can be made for such a function remains an open question. While working with causal sets, the Hadamard condition becomes a non-issue since there is natural discreteness and the question of UV limiting behavior does not arise.  

\subsection{SJ in the Causal Set} 
\label{sec:numerics} 

Causal sets are a natural covariant discretisation of the continuum, they also contain important signatures of quantum spacetime. This makes the results of simulations on causal sets interesting. As mentioned before, the SJ construction simplifies drastically when working with causal sets because the central eigenvalue equation for the PJ operator is now reduced to a matrix equation. Therefore, simulations are only limited by the size of the matrices involved, which is the same as the size of the causal set. 

Before we present examples, a quick dimensional analysis tells us the right quantities to compare -

The retarded Green function in the continuum satisfies the KG equation so\footnote{$[\,]$ refers to length dimension.} $[G]=2-d=[\Delta]=[W]$. Therefore $[\lambda_k]=2$ and the normalised eigenfunctions have $[\uksj]=1-d/2$.

In the causal set, we get the dimension of the massless Green function $K_0$ by requiring that $[K_0m^2/\rho]=0$, where $[m^2/\rho]=d-2$. This gives $[K_0]=2-d=[G]=[i\Delta]$. Using the correspondence $\int dV_y\rightarrow\frac{1}{\rho}\sum_y$, the eigenvalue equation becomes a matrix equation
\be \label{eq:csee}
\frac{1}{\rho}\,i\Delta f_k=\lambda_k\,f_k\,.
\ee   
where\footnote{In simulations the $1/\rho$ factor in Eq.\eqref{eq:csee} is omitted, which is why in the figures showing the eigenvalues are divided by $\rho$.} $[\lambda_k]=2$. 

As in the continuum, we have\footnote{Here, normalisation is obtained by taking the dot product of the vector with itself, divided by the density.} $[\uksj]=1-d/2$. Further, $[W]=2-d$ and can be compared directly with its counterpart in the continuum.

We show numerical results for the causal set SJ vacuum for 4 cases \cite{Afshordi:2012ez,Surya:2018byh}- causal diamonds in 2d and 4d Minkowski spacetime and slabs of 2d and 4d global de Sitter spacetime. In all cases we show how the spectrum of the PJ operator compares between the continuum and the causal set. We also show the SJ vacuum and its comparison with the continuum, whenever applicable. Note that, where visible, error bars in the binned data reflect the standard error of the mean (SEM).

\subsubsection{Causal Diamond in $\mink^2$}
 
The IR-regulated Minkowski two-point function is
\be \label{eq:massless2dmink}
	\text{Re}[\Wmink]=-\frac{1}{2 \pi} \ln(x) + c_1,\quad \quad x=\tau \,\text{or}\, d, 
\ee

where $\tau,d$ are the timelike, spacelike distances respectively and $c_1$ depends on the IR cutoff. In \cite{Afshordi:2012ez} it was shown that in a small subregion in the center of the causal diamond (i.e., away from the boundaries)

\be
	c_1\approx -\frac{1}{2 \pi} \ln(\lambda e^\gamma),
\ee
where $\gamma$ is the Euler-Mascheroni constant and $\lambda\sim0.46/L$, and where $2L$ is the side length of the diamond. In units where the volume (area in 2$d$) of the diamond is unity, $L=1/2$ and $c_1\approx -0.0786$.  

Simulation results are shown in figures \ref{2dcdeigens}-\ref{2dcdsub}. Figure \ref{2dcdeigens} is a log-log plot of the positive causal set SJ eigenvalues, along with the positive continuum eigenvalues - the two sets of eigenvalues are in agreement up to a characteristic ``knee" at which the causal set spectrum dips and ceases to obey a power-law with exponent $-1$. Specifically, the causal set SJ eigenvalues can be modelled as $\lambda^{\text{CS}}=\beta_1/n^\alpha+\beta_2\,n+\beta_3$. Importantly, there is a smooth, linear regime in the UV and this non-scaling behavior could indicate new physics \cite{Mathur:2022ivs}. There is also a clear convergence of the spectrum with causal set size $N$ except that the knee is pushed to smaller eigenvalues as $N$ increases. We see these patterns in all cases that we consider.  

\bfig[H]
\begin{center} 
	\includegraphics[width = .7\textwidth]{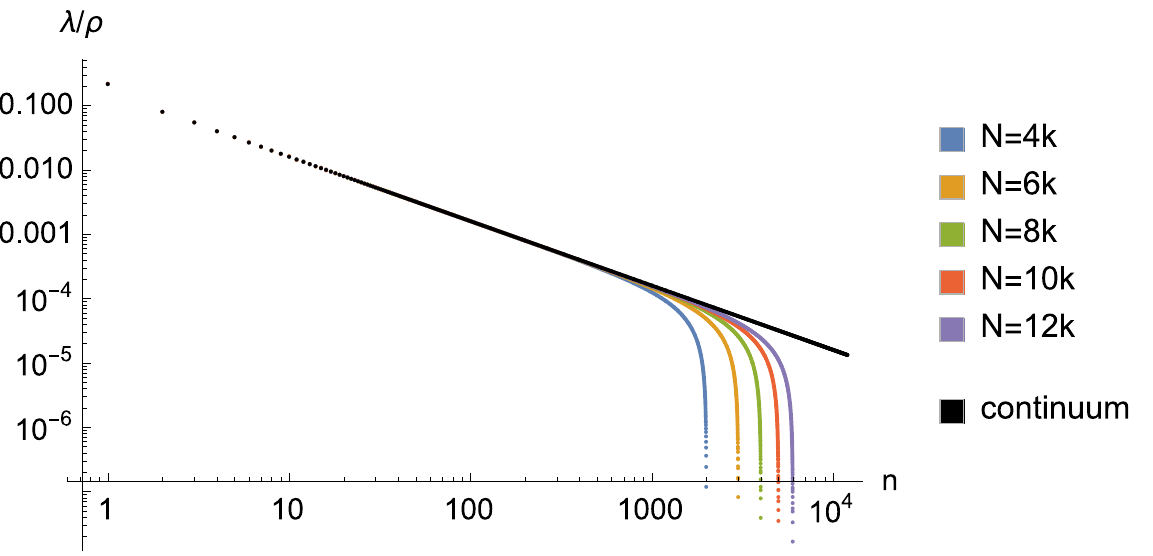}
	\caption{Log-log plot of the eigenvalues of $i\Delta$ divided by density $\rho$ (for the causal sets, the continuum spectrum is the straight black line), in the 2d causal diamond; $m=0$.}
	\label{2dcdeigens}
\end{center} 
\efig
Figure \ref{2dcd} shows scatter plots of $\text{Re}[\wsj]$ for pairs of events that are causally and spacelike related; it also shows the binned and averaged plots where the convergence becomes clear. The convergence with $N$ is very good and tells us that we are in the asymptotic regime i.e., $N$ is large enough. A comparison with the continuum is shown in figure \ref{2dcdsub}, it shows the scatter plots and the binned and averaged plots for $\wsj$ within a smaller diamond of side length $1/4$ compared to that of the original diamond it is concentric to. The continuum IR-regulated Minkowski curve is also plotted. These plots confirm that away from the boundaries of the diamond $\tre[\wsj]$ indeed resembles the Minkowski vacuum.  
\bfig[H]
\bsfig[b]{0.4\textwidth}
\includegraphics[width=\textwidth]{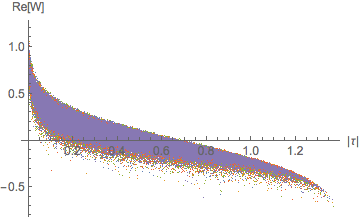}
\caption{Causal}
\esfig
\bsfig[b]{0.4\textwidth}
\includegraphics[width=\textwidth]{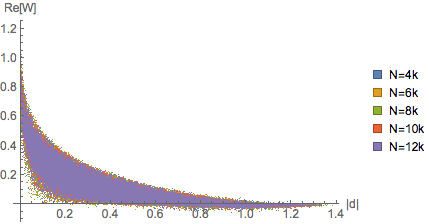}
\caption{Spacelike}
\esfig\\
\bsfig[b]{0.4\textwidth}
\includegraphics[width=\textwidth]{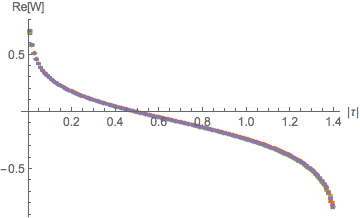}
\caption{Causal}
\esfig
\bsfig[b]{0.4\textwidth}
\includegraphics[width=\textwidth]{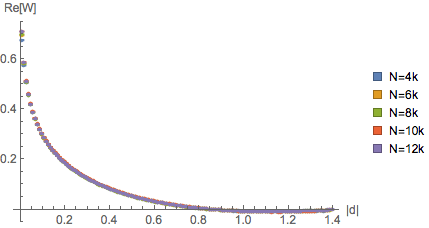}
\caption{Spacelike}
\esfig
\caption{(a)-(b) represent $\tre[\wsj]$ vs. geodesic distance for a sample of $100000$ randomly selected pairs, in the 2d causal diamond; $m=0$. (c)-(d) are plots of the binned and averaged data with the SEM.}
\label{2dcd} 
\efig

\bfig[H]
\bsfig[b]{0.4\textwidth}
\includegraphics[width=\textwidth]{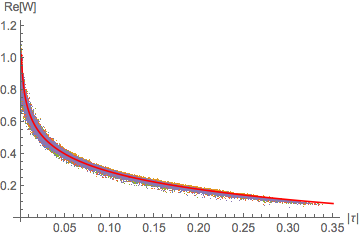}
\caption{Causal}
\esfig
\bsfig[b]{0.4\textwidth}
\includegraphics[width=\textwidth]{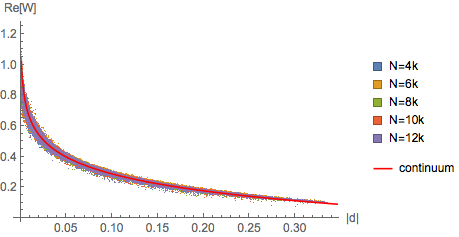}
\caption{Spacelike}
\esfig\\
\bsfig[b]{0.4\textwidth}
\includegraphics[width=\textwidth]{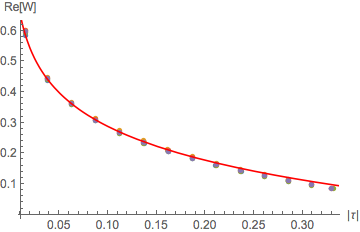}
\caption{Causal}
\esfig
\bsfig[b]{0.4\textwidth}
\includegraphics[width=\textwidth]{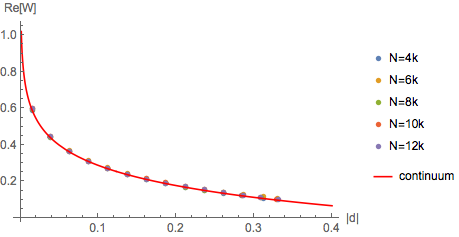}
\caption{Spacelike}
\esfig
\caption{(a)-(b) represent $\tre[\wsj]$ vs. geodesic distance for all pairs within a sub-diamond with side length $1/4$ of that the full diamond, in the 2d causal diamond; $m=0$. (c)-(d) are plots of the binned and averaged data with the SEM. In both cases, the continuum IR-regulated Minkowski Wightman function Eq.\eqref{eq:massless2dmink} is also shown.}
\label{2dcdsub}
\efig

\subsubsection{ Causal Diamond in $\mink^4$} 

The 4d Minkowski two-point function is 
\be \label{eq:4dmink}
	\text{Re}[\Wmink]=\frac{1}{4 \pi^2 x^2},\quad x=i\tau \,\text{or}\, |d| ,
\ee
and we work in units where the height of the diamond is unity.

In figure \ref{4dcdeigens} we show the log-log plot of the SJ spectrum which converges well as $N$ is increased, except near the knee which, as in the 2d diamond, shifts to the UV as $N$ increases. Also, there is no analytic calculation of the SJ spectrum in this case to compare with. 

\bfig[H]
\begin{center} 
	\includegraphics[width = .6 \textwidth]{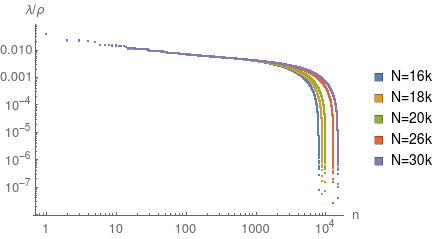}
	\caption{Log-log plot of the eigenvalues of $i\Delta$ divided by density $\rho$, in the 4d causal diamond; $m=0$.}
 \label{4dcdeigens}
\end{center} 
\efig
In figure \ref{4dcd} we show the scatter and binned plots for $\text{Re}[\wsj]$ as $N$ is varied. The convergence with increasing density suggests that the larger $N$ values are approaching the asymptotic regime. The Minkowski two-point function Eq.\eqref{eq:4dmink} is also included in this plot and it clearly does not agree with $\wsj$ in the full diamond. The small distance behaviour shows departure from the continuum, softening the divergences. 

\bfig[H] 
\bsfig[b]{0.46\textwidth}
\includegraphics[width=\textwidth]{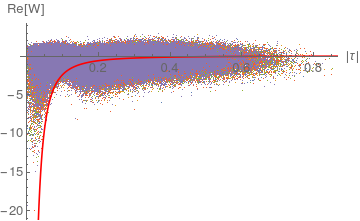}
\caption{Causal}
\esfig
\bsfig[b]{0.53\textwidth}
\includegraphics[width=\textwidth]{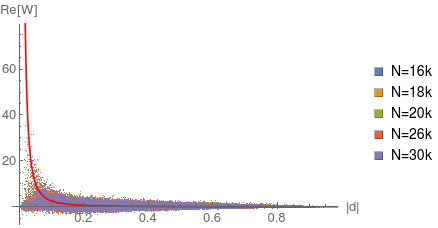}
\caption{Spacelike}
\esfig\\
\bsfig[b]{0.46\textwidth}
\includegraphics[width=\textwidth]{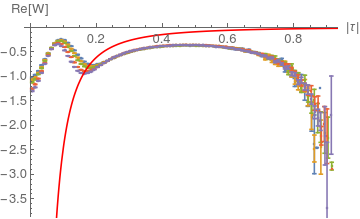}
\caption{Causal}
\esfig
\bsfig[b]{0.53\textwidth}
\includegraphics[width=\textwidth]{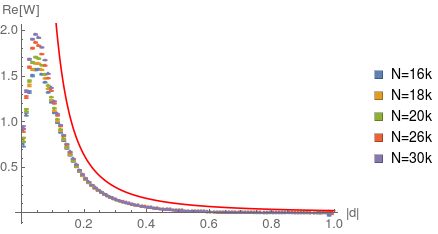}
\caption{Spacelike}
\esfig
\caption{(a)-(b) represent $\tre[\wsj]$ vs. geodesic distance for a sample of $100000$ randomly selected pairs, in the 4d causal diamond; $m=0$. (c)-(d) are plots of the binned and averaged data with the SEM. In both cases, the continuum Minkowski Wightman function Eq.\eqref{eq:4dmink} is shown in red.}
\label{4dcd}
\efig
Figure \ref{4dcdsub} shows the scatter and binned plots for a smaller causal diamond of side length $1/2$ compared to the larger diamond it is in the center of. Although the agreement of $\wsj$ with $\Wmink$ is not as good as in 2d, we see that as $N$ increases, there is a convergence of $\wsj$ to $\Wmink$. This suggests that as in 2d, the 4d diamond also shows an agreement with the Minkowski vacuum far away from the boundary. 
 
\bfig[H]
\bsfig[b]{0.45\textwidth}
\includegraphics[width=\textwidth]{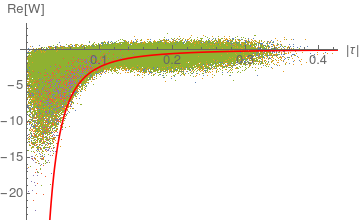}
\caption{Causal}
\esfig
\bsfig[b]{0.54\textwidth}
\includegraphics[width=\textwidth]{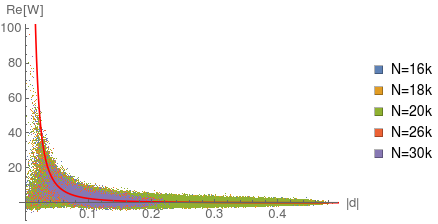}
\caption{Spacelike}
\esfig\\
\bsfig[b]{0.45\textwidth}
\includegraphics[width=\textwidth]{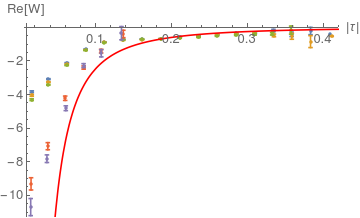}
\caption{Causal}
\esfig
\bsfig[b]{0.54\textwidth}
\includegraphics[width=\textwidth]{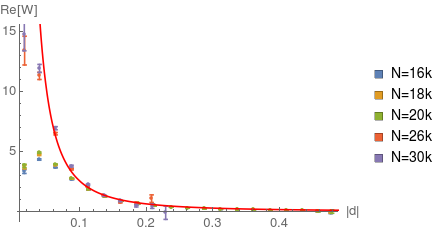}
\caption{Spacelike}
\esfig
\caption{(a)-(b) represent $\tre[\wsj]$ vs. geodesic distance for all pairs within a sub-diamond with height $1/2$ of the full diamond, in the 4d causal diamond; $m=0$. (c)-(d) are plots of the binned and averaged data with the SEM. In both cases, the continuum Minkowski Wightman function Eq.\eqref{eq:4dmink} is also shown.}
\label{4dcdsub}
\efig

\subsubsection{Slab in dS$^2$}

The slab we consider in de Sitter spacetime lies within the region $[-T,T]$\footnote{$T$ is the cutoff in the conformal time.}. We also need to check convergence with $T$ at fixed $\rho$, to show that the results are independent of the IR cutoff.  

The Wightman function for the Euclidean vacuum in $d$ spacetime dimensions is given by\footnote{The expression for $W_E$ in equation $B.36$ of \cite{Aslanbeigi:2013fga} has a minor typographical error: the factor of $4\pi$ should be raised to the power of $d/2$. See for example \cite{Bousso2002conformal}.}
\be \label{eq:we}
W_E(x,y)=\frac{\Gamma[h_+] \Gamma[h_-]}{(4 \pi)^{d/2}\ell^2 \Gamma[\frac{d}{2}]}\,  {}_2F_1\left(h_+,h_-, \frac{d}{2};\frac{1+Z(x,y)+i\epsilon\, \text{sign}(x^0-y^0)}{2}\right),
\ee
where $Z(x, y)$ is related to the geodesic distance,  $h_\pm=\frac{d-1}{2}\pm\nu$, $\nu=\ell\sqrt{m_*^2-m^2}$, $m_*=\frac{d-1}{2\ell}$ and $_2F_1(a,b,c;z)$ is a hypergeometric function. The symmetric two-point function, or Hadamard function, for any other Allen-Mottola $\alpha$-vacuum is \cite{Aslanbeigi:2013fga}
\be \label{eq:walpha}
H_{\alpha\beta}(x,x')=\cosh2\alpha\,H_E(x,x')+\sinh2\alpha\,[\cos\beta\,H_E(\bar{x},x')-\sin\beta\,\Delta(\bar{x},x')],
\ee
where $\bar{x}$ is the antipodal point of $x$. The Wightman function is related to $H$ by $2W=H+i\Delta$. We show comparisons with the $\alpha$-vacua found to correspond to the SJ vacuum in \cite{Aslanbeigi:2013fga}. Since we work in even dimensions, these are
$\alpha=0$ for $m\geq m_*$ (yielding the Euclidean vacuum), and 
\be
\alpha=\frac{1}{2}\tanh^{-1}|\sin\pi\nu|\quad\text{and}\quad\beta=\pi[\frac{d}{2}+\theta(-\sin\pi\nu)]
\ee
for $m<m_*$.

We begin with 2d de Sitter spacetime, and work in units in which the de Sitter radius $\ell=1$. In 2d,  $m_*= 0.5$, and the conformal mass $m_c=0$. Hence the minimally coupled and the conformally coupled massless cases coincide.
The simulations span slabs of different $T$ values from $1$ to $1.5$, while $N$ values range from $8k$ to $36k$. We show the log-log plots of the PJ spectrum for the massless $m=0$ and for the massive\footnote{This is an arbitrary choice of mass with no special physical significance. It  allows for comparisons with \cite{Aslanbeigi:2013fga} in their 2d de Sitter causal set simulations.} $m=2.3$ cases in figure \ref{2ddSspectrum}. The spectrum converges well for both masses, with the knee shifting to the UV as $N$ increases, as expected. We also show the comparison between the causal set spectrum with the finite $T$ continuum spectrum obtained via the mode comparison method in \cite{Aslanbeigi:2013fga}. As shown in figure \ref{2ddSspectrum} this spectrum does not seem to agree with the causal set spectrum even though the latter convergences with $N$. 

\bfig[H]
\bsfig[b]{0.43\textwidth}
\includegraphics[width=\textwidth]{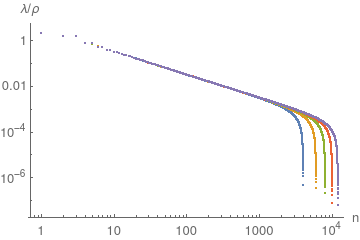}
\caption{$m=0$}
\esfig
\bsfig[b]{0.57\textwidth}
\includegraphics[width=\textwidth]{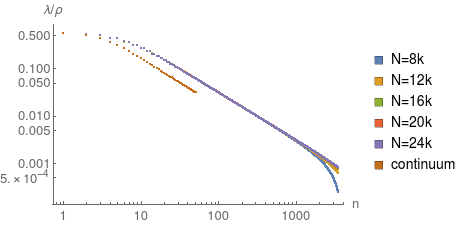}
\caption{$m=2.3$}
\esfig
\caption{Log-log plot of the positive eigenvalues of $i\Delta$ at $T=1$, in 2d de Sitter. In the massive case on the right we plot the largest 3500 positive eigenvalues and the corresponding continuum eigenvalues from the finite T mode comparison results of \cite{Aslanbeigi:2013fga}.} 
\label{2ddSspectrum}
\efig

We show $\wsj$ for both the above masses: $m=0$ and $m=2.3$, and vary over both the slab height $T$ and density $\rho$. For $m=2.3$, as can be seen in the scatter plots of figures \ref{2ddSSJ1}, \ref{2ddSSJ15} and \ref{2ddST156}, $\wsj$ agrees very well with the SJ vacuum expected from the calculation in \cite{Aslanbeigi:2013fga} (the Euclidean vacuum). Furthermore, it appears that $\wsj$ for a given $T$ is simply the restriction of $\wsj$ for a larger $T$. This is also in agreement with \cite{Aslanbeigi:2013fga}. 

For the massless case, the scatter plots of $\wsj$ in figures \ref{2ddSSJ1zero}, \ref{2ddSSJ15zero} and \ref{2ddST156zero} do not show convergence, but instead fan out, as a function of the proper time and distance. As the density decreases, for $T=1.56, N=36k$, the scatter plot figure \ref{2ddST156zero} shows a clustering into two distinct sets. This suggests that $\wsj$ may not just be a function of just proper time and distance, and hence may not be de Sitter invariant. 

\bfig[H]  
\bsfig[b]{0.42\textwidth}
\includegraphics[width=\textwidth]{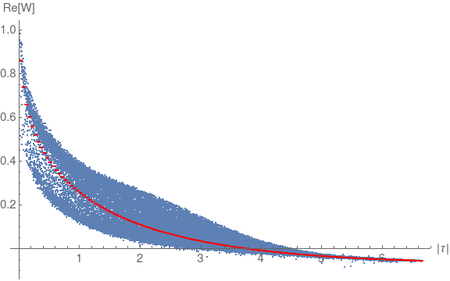}
\caption{Causal $m=0$}
\esfig
\bsfig[b]{0.42\textwidth}
\includegraphics[width=\textwidth]{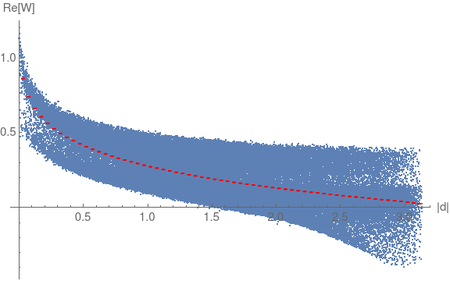}
\caption{Spacelike $m=0$}
\esfig
\caption{$N=32000, T=1, \rho=1635.08$, in 2d de Sitter. (a)-(b) represent $\tre[\wsj]$ vs. geodesic distance for a sample of 100000 randomly selected pairs, and the red curve represents the mean values with the SEM.}
\label{2ddSSJ1zero}
\efig

\bfig[H]
\bsfig[b]{0.45\textwidth}
\includegraphics[width=\textwidth]{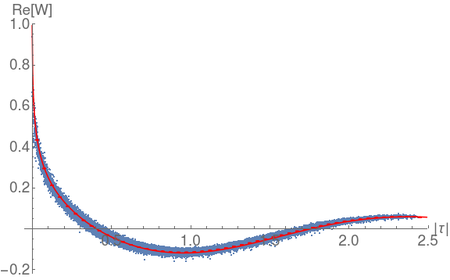}
\caption{Causal $m=2.3$}
\esfig
\bsfig[b]{0.45\textwidth}
\includegraphics[width=\textwidth]{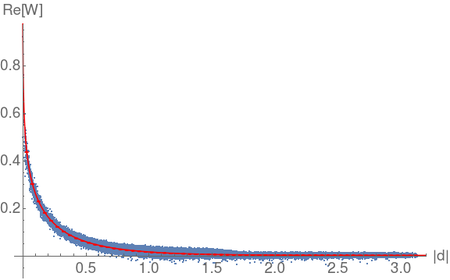}
\caption{Spacelike $m=2.3$}
\esfig
\caption{$N=24000,\, T=1,\, \rho=1226.31$, in 2d de Sitter. The scatter plot is $\tre[\wsj]$ vs. geodesic distance for a sample of 100000 randomly selected pairs. The red curve represents the continuum $W_E$ from Eq.\eqref{eq:we}.}
\label{2ddSSJ1} 
\efig
\bfig[H]
\bsfig[b]{0.45\textwidth}
\includegraphics[width=\textwidth]{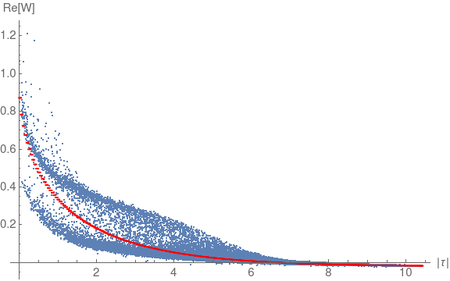}
\caption{Causal $m=0$}
\esfig
\bsfig[b]{0.45\textwidth}
\includegraphics[width=\textwidth]{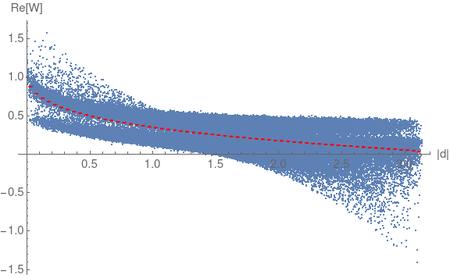}
\caption{Spacelike $m=0$}
\esfig
\caption{$N=36000, T=1.5, \rho=203.15$, in 2d de Sitter. (a)-(b) represent $\tre[\wsj]$ vs. geodesic distance for a sample of 100000 randomly selected pairs. The red curve represents the mean values with the SEM.}
\label{2ddSSJ15zero} 
\efig
\bfig[H]
\bsfig[b]{0.45\textwidth}
\includegraphics[width=\textwidth]{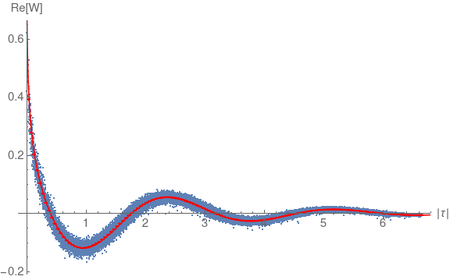}
\caption{Causal $m=2.3$}
\esfig
\bsfig[b]{0.45\textwidth}
\includegraphics[width=\textwidth]{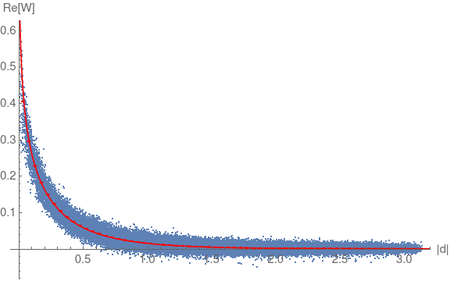}
\caption{Spacelike $m=2.3$}
\esfig
\caption{$N=36000,\, T=1.5,\, \rho=203.15$, in 2d de Sitter.  $\tre[\wsj]$ vs. geodesic distance for 100000 randomly selected pairs. The red curve represents the continuum $W_E$ from Eq.\eqref{eq:we}.}
\label{2ddSSJ15} 
\efig
\bfig[H]
\bsfig[b]{0.45\textwidth}
\includegraphics[width=\textwidth]{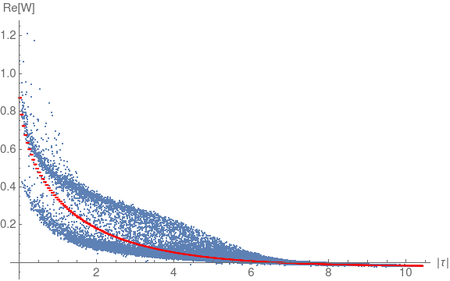}
\caption{Causal $m=0$}
\esfig
\bsfig[b]{0.45\textwidth}
\includegraphics[width=\textwidth]{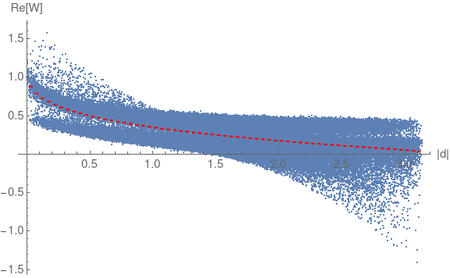}
\caption{Spacelike $m=0$}
\esfig
\caption{$N=36000, -1.56<\tilde{T}<1.56, \rho=30.93$, in 2d de Sitter. (a)-(b) represent $\tre[\wsj]$ vs. geodesic distance for a sample of  100000 randomly selected pairs. The red curve represents the mean values with the SEM.}
\label{2ddST156zero} 
\efig
\bfig[H]
\bsfig[b]{0.45\textwidth}
\includegraphics[width=\textwidth]{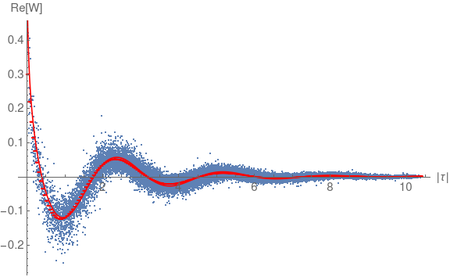}
\caption{Causal $m=2.3$}
\esfig
\bsfig[b]{0.45\textwidth}
\includegraphics[width=\textwidth]{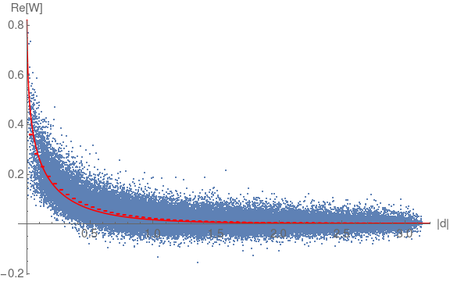}
\caption{Spacelike $m=2.3$}
\esfig
\caption{$N=36000, T=1.56, \rho=30.93$, in 2d de Sitter. $\tre[\wsj]$ vs. geodesic distance for a sample of 100000 randomly selected pairs. The red curve represents the continuum $W_E$ from Eq.\eqref{eq:we}.}
\label{2ddST156} 
\efig

\subsubsection{Slab in dS$^4$}

Finally, we present simulations for the 4d de Sitter SJ vacuum. In 4d, $m_*=1.5$ and $m_c=\sqrt{2}\approx 1.41$. 

Figure \ref{4ddSeigen} shows the log-log plot of the SJ spectrum for $m=0$ and $m=2.3$ for various $N$. We find excellent convergence with $N$ in both cases, and again, as in the other cases, there is a knee which shifts to the UV as $N$ is increased. However, there is poor agreement with the continuum values of the finite $T$ spectrum calculated via the mode comparison method in \cite{Aslanbeigi:2013fga}, as in the 2d case. There is also no unusual behaviour close to the masses $m=0$ and $m=m_c\approx1.41$ \cite{Surya:2018byh}.    

\bfig[H]
\bsfig[b]{0.46\textwidth}
\includegraphics[width=\textwidth]{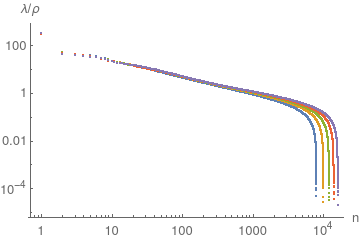}
\caption{$m=0$}
\esfig
\bsfig[b]{0.53\textwidth}
\includegraphics[width=\textwidth]{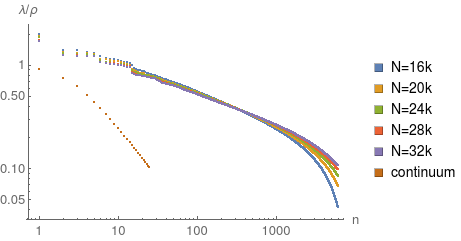}
\caption{$m=2.3$}
\esfig
\caption{Log-log plot of the positive eigenvalues of $i\Delta$, in 4d de Sitter. In the massive case on the right we plot the largest 6000 positive eigenvalues and the corresponding continuum eigenvalues from the finite T mode comparison results of \cite{Aslanbeigi:2013fga}.}
\label{4ddSeigen}
\efig

Figures \ref{4ddSzero} and \ref{4ddS} are sample scatter plots of $\wsj$ for $m=0$ and $m=2.3$.
In figure \ref{4ddSfixedT} we fix $T$ for $m=0$ and for $m=m_c\approx1.41$ and vary $N$ to check for convergence with density; for smaller proper times and distances, the convergence is not as good as it is for larger proper times and distances. For $m=1.41$ we also plot the  Wightman function associated with the  Euclidean vacuum $W_E$ in Eq.\eqref{eq:we}. $W_E$ does not compare well with the causal set $\wsj$. 
The convergence with $T$ can also be checked and is good for various $m$ values. However, the Wightman function associated with the $\alpha$-vacuum Eq.\eqref{eq:walpha} as well as the Euclidean vacuum $W_E$ once again do not compare well with the causal set $\wsj$ for any of these masses. Overall, simulations strongly suggest that the causal set 4d de Sitter $\wsj$ differs from the Mottola-Allen $\alpha$-vacua for all masses. In particular, there is modification of the small distance behaviour of the state, making it well defined in the UV limit.  

\bfig[H]
\bsfig[b]{0.48\textwidth}
\includegraphics[width=\textwidth]{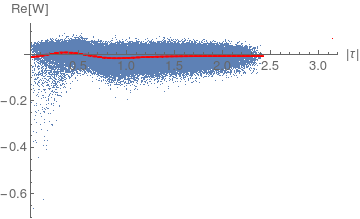}
\caption{Causal $T=1$}
\esfig
\bsfig[b]{0.48\textwidth}
\includegraphics[width=\textwidth]{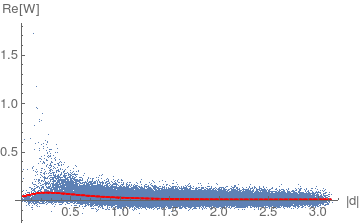}
\caption{Spacelike $T=1$}
\esfig\\
\bsfig[b]{0.48\textwidth}
\includegraphics[width=\textwidth]{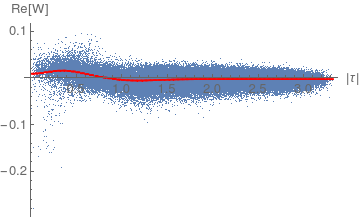}
\caption{Causal $T=1.2$}
\esfig
\bsfig[b]{0.48\textwidth}
\includegraphics[width=\textwidth]{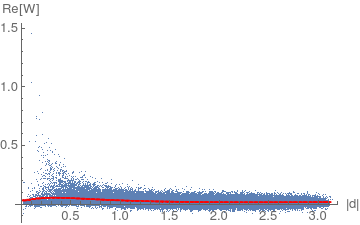}
\caption{Spacelike $T=1.2$}
\esfig
\caption{$m=0,\,N=32000$, in 4d de Sitter. $\tre[\wsj]$ vs. geodesic distance for   100000 randomly selected pairs,  and the red curve represents the mean
	values with the SEM.}
\label{4ddSzero}
\efig

\bfig[H]
\bsfig[b]{0.48\textwidth}
\includegraphics[width=\textwidth]{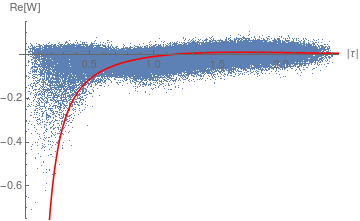}
\caption{Causal $T=1$}
\esfig
\bsfig[b]{0.48\textwidth}
\includegraphics[width=\textwidth]{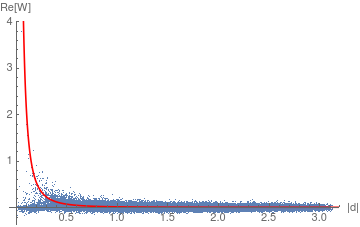}
\caption{Spacelike $T=1$}
\esfig\\
\bsfig[b]{0.48\textwidth}
\includegraphics[width=\textwidth]{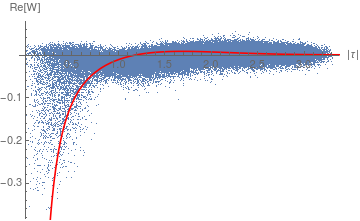}
\caption{Causal $T=1.2$}
\esfig
\bsfig[b]{0.48\textwidth}
\includegraphics[width=\textwidth]{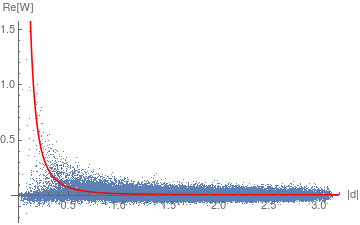}
\caption{Spacelike $T=1.2$}
\esfig
\caption{$m=2.3,\,N=32000$, in 4d de Sitter. $\tre[\wsj]$ vs. geodesic distance for a sample of  100000 randomly selected pairs. The red curve shows the Euclidean two-point function $W_E$ from Eq.\eqref{eq:we}.} 
\label{4ddS}
\efig
 
\bfig[H]
\bsfig[b]{0.46\textwidth}
\includegraphics[width=\textwidth]{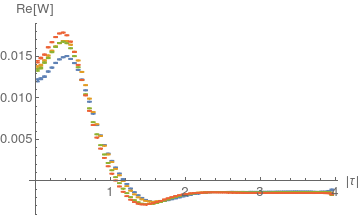}
\caption{Causal $m=0,T=1.3$}
\esfig
\bsfig[b]{0.53\textwidth}
\includegraphics[width=\textwidth]{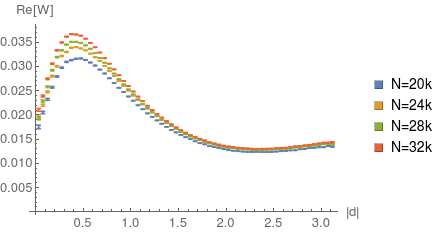}
\caption{Spacelike $m=0,T=1.3$}
\esfig\\
\bsfig[b]{0.46\textwidth}
\includegraphics[width=\textwidth]{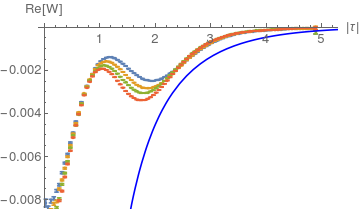}
\caption{Causal $m=1.41,T=1.4$}
\esfig
\bsfig[b]{0.53\textwidth}
\includegraphics[width=\textwidth]{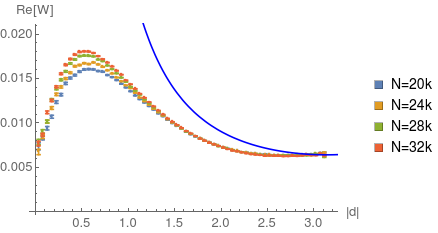}
\caption{Spacelike $m=1.41,T=1.4$}
\esfig
\caption{$\tre[\wsj]$ vs. geodesic distance with varying density, in 4d de Sitter. The  blue curve shows the Euclidean two-point function as a reference.}
\label{4ddSfixedT}
\efig

\section{Concluding Remarks}

Before concluding, We mention two ideas that have been explored around building a QFT on causal sets - 

\begin{itemize}

\item \textit{Equations of motion}: This arises as an aside to the SJ construction is the causal set - namely looking at time evolution as a set of algebraic constraints on the field configaration rather than a discretized differential operator (i.e., the d'Alembertian). Such constraints can be obtained from the relation 
$$
\kr (\hB -\phm^2) = \overline{\im (\hD)}.
$$
In principle this is applicable to a globally hyperbolic region of any general spacetime.  

On the causal set, a solution of the field equation $b\in\kr (\hB)$ also lies in $\im (\hD)$, therefore,\footnote{We use indices here instead of coordinates to emphasize that the objects here are matrices.}
\be
b_x=\sum_{k=1}^r a_k s_{kx}
\ee   
where $s_{kx}$ span $\im (\hD)$ are just the SJ modes and $r=\text{dim}(\im (\hD))$. We can think of this equation as an initial value problem by assigning known values to some "initial points" in the causal set i.e., some values in the solution vector $b$ can be assigned as initial values. A preliminary discussion on this can be found in \cite{Nomaan2021}.       

\item \textit{Fermions}: To discuss QFT for fermions, we need a way to first obtain the retarted Green function analogue on the causal set and then use an appropriate PJ operator in order to get a unique set of modes. The latter issue is resolved easily by replacing the bosonic PJ operator, which is written as a commutator of field operators, with a fermionic version based on an anti-commutator, i.e.,
$$
\{\hat{\phi}_{\alpha}(x),\hat{\phi}_{\beta}(y)^\dagger\}=i\Delta_{\alpha,\beta}(x,y).
$$
Here $\alpha,\beta$ are spin indices. 

In \cite{Johnston:2010su}, Johnston gives two proposals for constructing spin-1/2 Green functions on the causal set. The first one uses the Feynmann checkerboard idea \cite{feynman2010quantum} which involves a path sum involving paths that zig-zag in spacetime and are made up of null geodesics. The second one uses the notion that the Green function for the Dirac equation can be constructed as a ``square root" of the Klein-Gordon Green function. The main insight here is that if we define 
$$
S_m(x-y)\equiv-\int\,d^dz\,R_m(x-z)R_{-m}(z-y)
$$
such that $R$ satisfies the dirac equation, then $S$ satisfies the Klein-Gordon equation. Like in the bosonic case, we can use a series expansion to get $R_m$ once $R_0$ is determined. A similar construction can be used on causal sets. However, the square root of a matrix (if it exists) is not unique in general and therefore we may need further constraints. 

\end{itemize}

QFT on causal sets is an important direction in the broad area of computational (numerical) QFT and in discrete approaches to quantum gravity. While we are not working here in the deep UV regime of full quantum gravity, we are working at a mesoscale where the effects of discreteness are still relevant to the dynamics of the quantum field. We showed how a free scalar field theory can be set up in various situations through the use of the SJ method. We also mentioned other proposals around describing matter on a causal set. Such studies still have a long way to go. Firstly, there are conceptual implications of of defining such a state - we mentioned that the SJ state coincides with the usual vacuum in static spacetimes. This is a limited result and it is not always clear how the SJ state corresponds to other states that can be defined in general spacetimes. Mathematically, this is a consequence of working with finite regions, while other standard states are usually defined in full spacetime. The answer to the question - "the SJ state is the natural state for which observer?" is not obvious. Secondly, we would like to apply this construction to more realistic theories - interacting theories and theories involving fermions. The true test of utility and of phenomenological relevance has to come from S-matrix calculations for these theories. Some exciting ideas have been proposed in these directions \cite{Johnston:2010su,Hawkins2022} and these need to be pushed further in order to deepen our understanding of the propagation of quantum fields on causal sets. 
\bibliography{refs_qftcs}{}

\begin{thebibliography}{10}
\providecommand{\url}[1]{{#1}}
\providecommand{\urlprefix}{URL }
\expandafter\ifx\csname urlstyle\endcsname\relax
  \providecommand{\doi}[1]{DOI~\discretionary{}{}{}#1}\else
  \providecommand{\doi}{DOI~\discretionary{}{}{}\begingroup
  \urlstyle{rm}\Url}\fi

\bibitem{Afshordi:2012jf}
Afshordi, N., Aslanbeigi, S., Sorkin, R.D.: {A Distinguished Vacuum State for a
  Quantum Field in a Curved Spacetime: Formalism, Features, and Cosmology}.
\newblock JHEP \textbf{08}, 137 (2012).
\newblock \doi{10.1007/JHEP08(2012)137}

\bibitem{Afshordi:2012ez}
Afshordi, N., Buck, M., Dowker, F., Rideout, D., Sorkin, R.D., Yazdi, Y.K.: {A
  Ground State for the Causal Diamond in 2 Dimensions}.
\newblock JHEP \textbf{10}, 088 (2012).
\newblock \doi{10.1007/JHEP10(2012)088}

\bibitem{Aslanbeigi:2013fga}
Aslanbeigi, S., Buck, M.: {A preferred ground state for the scalar field in
  deSitter space}.
\newblock JHEP \textbf{08}, 039 (2013).
\newblock \doi{10.1007/JHEP08(2013)039}

\bibitem{aslanbeigi2014generalized}
Aslanbeigi, S., Saravani, M., Sorkin, R.D.: Generalized causal set
  d'alembertians.
\newblock Journal of High Energy Physics \textbf{2014}(6) (2014).
\newblock \doi{10.1007/jhep06(2014)024}

\bibitem{Aste:2007ii}
Aste, A.: {Resummation of mass terms in perturbative massless quantum field
  theory}.
\newblock Lett. Math. Phys. \textbf{81}, 77--92 (2007).
\newblock \doi{10.1007/s11005-007-0169-8}

\bibitem{Benincasa:2010ac}
Benincasa, D.M.T., Dowker, F.: {The Scalar Curvature of a Causal Set}.
\newblock Phys. Rev. Lett. \textbf{104}, 181301 (2010).
\newblock \doi{10.1103/PhysRevLett.104.181301}

\bibitem{Bousso2002conformal}
Bousso, R., Maloney, A., Strominger, A.: {Conformal vacua and entropy in de
  Sitter space}.
\newblock Phys. Rev. \textbf{D65}, 104039 (2002).
\newblock \doi{10.1103/PhysRevD.65.104039}

\bibitem{Brum:2013bia}
Brum, M., Fredenhagen, K.: {‘Vacuum-like’ Hadamard states for quantum
  fields on curved spacetimes}.
\newblock Class. Quant. Grav. \textbf{31}, 025024 (2014).
\newblock \doi{10.1088/0264-9381/31/2/025024}

\bibitem{Dable-Heath:2019sej}
Dable-Heath, E., Fewster, C.J., Rejzner, K., Woods, N.: {Algebraic Classical
  and Quantum Field Theory on Causal Sets}.
\newblock Phys. Rev. D \textbf{101}(6), 065013 (2020).
\newblock \doi{10.1103/PhysRevD.101.065013}

\bibitem{daughton1993the}
Daughton, A.R.: The recovery of locality for casual sets and related topics
  (1993)

\bibitem{dowker2013causal}
Dowker, F., Glaser, L.: Causal set d'alembertians for various dimensions.
\newblock Classical and Quantum Gravity \textbf{30}(19), 195016 (2013)

\bibitem{Fewster:2018ltq}
Fewster, C.J.: {The art of the state}.
\newblock Int. J. Mod. Phys. \textbf{D27}(11), 1843007 (2018).
\newblock \doi{10.1142/S0218271818430071}

\bibitem{fewster2012on}
Fewster, C.J., Verch, R.: {On a Recent Construction of 'Vacuum-like' Quantum
  Field States in Curved Spacetime}.
\newblock Class. Quant. Grav. \textbf{29}, 205017 (2012).
\newblock \doi{10.1088/0264-9381/29/20/205017}

\bibitem{feynman2010quantum}
Feynman, R.P., Hibbs, A.R., Styer, D.F.: Quantum mechanics and path integrals.
\newblock Courier Corporation (2010)

\bibitem{Foster_2004}
Foster, B.Z., Jacobson, T.: Quantum field theory on a growing lattice.
\newblock Journal of High Energy Physics \textbf{2004}(08), 024--024 (2004).
\newblock \doi{10.1088/1126-6708/2004/08/024}

\bibitem{Hawkins2022}
Hawkins, E., Minz, C., Rejzner, K.: Quantization, dequantization, and
  distinguished states  (2022)

\bibitem{Johnston2008particle}
Johnston, S.: {Particle propagators on discrete spacetime}.
\newblock Class. Quant. Grav. \textbf{25}, 202001 (2008).
\newblock \doi{10.1088/0264-9381/25/20/202001}

\bibitem{johnston2015correction}
{Johnston}, S.: Correction terms for propagators and d'alembertians due to
  spacetime discreteness.
\newblock Classical and Quantum Gravity \textbf{32}(19), 195020 (2015).
\newblock \doi{10.1088/0264-9381/32/19/195020}

\bibitem{Johnston:2010su}
Johnston, S.P.: {Quantum Fields on Causal Sets}.
\newblock Ph.D. thesis, Imperial Coll., London (2010).
\newblock
  \urlprefix\url{http://inspirehep.net/record/874679/files/arXiv:1010.5514.pdf}

\bibitem{Mathur2019}
Mathur, A., Surya, S.: {Sorkin-Johnston vacuum for a massive scalar field in
  the 2D causal diamond}.
\newblock Phys. Rev. \textbf{D100}(4), 045007 (2019).
\newblock \doi{10.1103/PhysRevD.100.045007}

\bibitem{Mathur:2022ivs}
Mathur, A., Surya, S., Nomaan, X.: {Spacetime entanglement entropy: covariance
  and discreteness}.
\newblock Gen. Rel. Grav. \textbf{54}(7), 74 (2022).
\newblock \doi{10.1007/s10714-022-02948-x}

\bibitem{Nomaan2021}
{Nomaan X}: Aspects of quantum fields on causal sets  (2021)

\bibitem{nomaan2017scalar}
{Nomaan X}, Dowker, F., Surya, S.: {Scalar Field Green Functions on Causal
  Sets}.
\newblock Class. Quant. Grav. \textbf{34}(12), 124002 (2017).
\newblock \doi{10.1088/1361-6382/aa6bc7}

\bibitem{salgado2008toward}
Salgado, R.B.: Toward a quantum dynamics for causal sets.
\newblock Syracuse University (2008)

\bibitem{sorkin2009does}
Sorkin, R.D.: Does locality fail at intermediate length-scales.
\newblock Approaches to quantum gravity: Toward a new understanding of space,
  time and matter pp. 26--43 (2009)

\bibitem{Sorkin:2011pn}
Sorkin, R.D.: {Scalar Field Theory on a Causal Set in Histories Form}.
\newblock J. Phys. Conf. Ser. \textbf{306}, 012017 (2011).
\newblock \doi{10.1088/1742-6596/306/1/012017}

\bibitem{Sorkin:2017fcp}
Sorkin, R.D.: {From Green Function to Quantum Field}.
\newblock Int. J. Geom. Meth. Mod. Phys. \textbf{14}(08), 1740007 (2017).
\newblock \doi{10.1142/S0219887817400072}

\bibitem{stanley1986enumerative}
Stanley, R.P.: Enumerative combinatorics, volume 1. wadsworth.
\newblock Inc. California  (1986)

\bibitem{Surya:2018byh}
Surya, S., {Nomaan X}, Yazdi, Y.K.: {Studies on the SJ Vacuum in de Sitter
  Spacetime}.
\newblock JHEP \textbf{07}, 009 (2019).
\newblock \doi{10.1007/JHEP07(2019)009}

\bibitem{sverdlov2009gravity}
{Sverdlov}, R., Bombelli, L.: Gravity and matter in causal set theory.
\newblock Classical and Quantum Gravity \textbf{26}(7), 075011 (2009)

\bibitem{Wald:1995yp}
Wald, R.M.: {Quantum Field Theory in Curved Space-Time and Black Hole
  Thermodynamics}.
\newblock Chicago Lectures in Physics. University of Chicago Press, Chicago, IL
  (1995)

\end{thebibliography}
\bibliographystyle{spmpsci} 
\end{document}